\documentclass[english,aps,prb,floats,twocolumn]{revtex4-2}
\usepackage[T1]{fontenc}
\usepackage[latin9]{inputenc}
\usepackage{amsmath}
\usepackage{amssymb}
\usepackage{graphicx}
\usepackage{wasysym}
\usepackage{esint}

\makeatletter
\usepackage{babel}

\makeatother

\usepackage{babel}
\begin{document}
\title{Scaling of Static and Dynamical Properties of Random Anisotropy Magnets}
\author{Dmitry A. Garanin and Eugene M. Chudnovsky}
\affiliation{Physics Department, Herbert H. Lehman College and Graduate School,
The City University of New York, 250 Bedford Park Boulevard West,
Bronx, New York 10468-1589, USA }
\date{\today}
\begin{abstract}
Recently observed scaling in the random-anisotropy model of amorphous
or sintered ferromagnets is derived by an alternative method and extended
for studying the dynamical properties in terms of the Landau-Lifshitz
equations for spin blocks. Switching to the rescaled exchange and
anisotropy constants allows one to investigate the dynamics by using
a reduced number of variables, which greatly speeds up computations.
The proposed dynamical scaling is applied to the problem of microwave
absorption by a random anisotropy magnet. The equivalence of the rescaled
model to the original atomic model is confirmed numerically. The method
is proposed as a powerful tool in studying static and dynamic properties
of systems with quenched randomness. 
\end{abstract}
\maketitle

\section{Introduction}

\label{Intro}

\begin{figure}[h]
\centering{}\includegraphics[width=8cm]{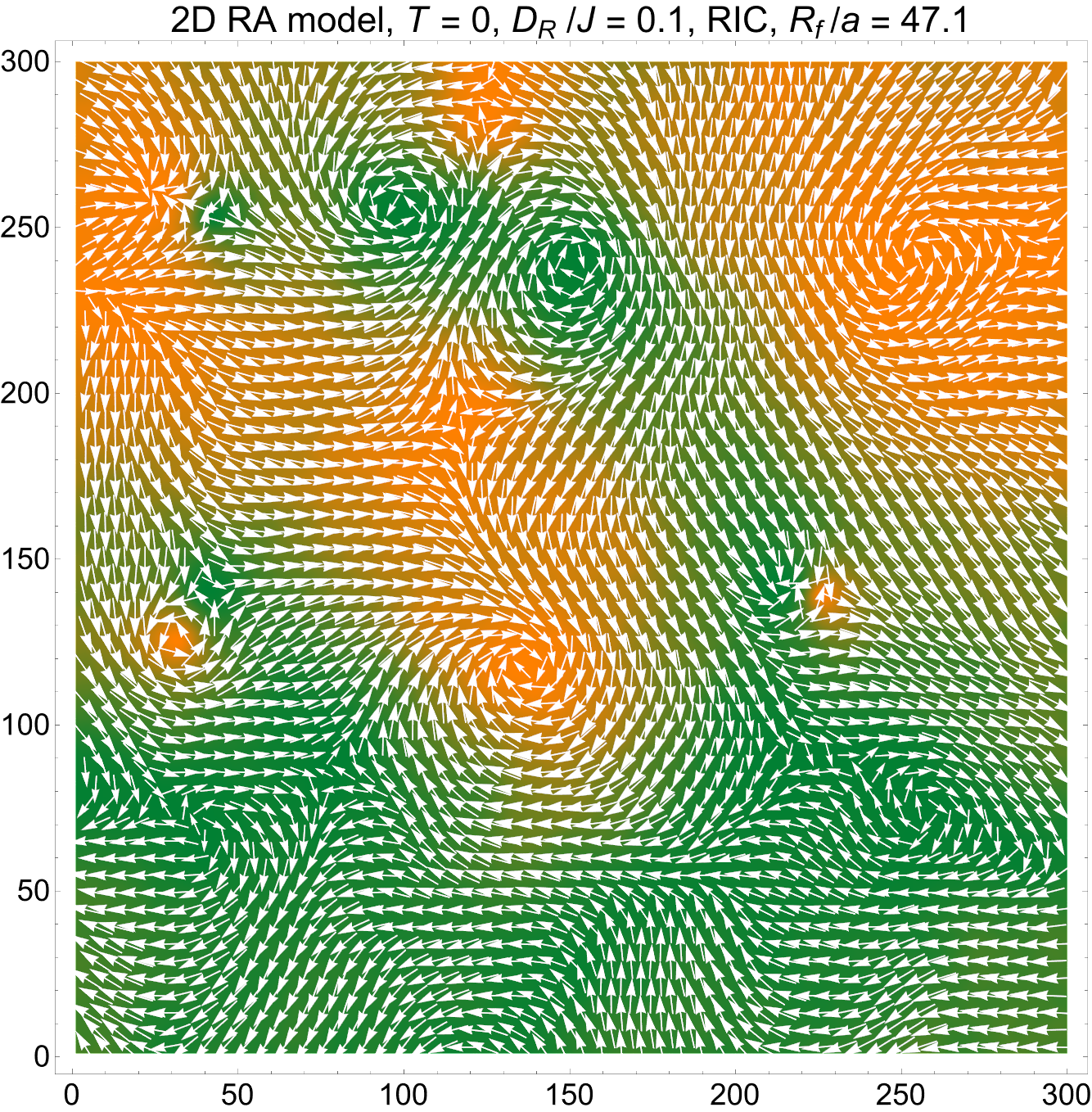}\caption{The equilibrium spin structure in a 2D system of spins with ferromagnetic
exchange interaction with free boundary conditions and random directions
of the magnetic anisotropy axes. In-plane spin components are shown
by white arrows. The out-of-plane component is shown by orange/green
corresponding to positive/negative. Except for isolated points representing
topological defects, the direction of local magnetization changes
on a scale $R_{f}$ large compared to the distance $a$ between neighboring
spins. }
\label{Fig-magnetic_structure} 
\end{figure}

All systems with a continuous order parameter possess some degree
of static randomness. Amorphous ferromagnets (see, e.g., Refs.\ \onlinecite{RA-book,CT-book,Marin-MMM2020,GC-JPhys2022}
and references therein) represent the extreme case of such randomness.
It destroys or strongly reduces the long-range ferromagnetic order
\citep{IM,CSS-1986,PCG-2015} and converts the ferromagnetic resonance
(FMR) into a broad absorption band even at zero temperature \citep{GC-PRB2021}.
Qualitatively, these effects are caught by the random anisotropy (RA)
model \citep{Harris-PRL1973} which considers random directions of
magnetic anisotropy axes in the presence of the ferromagnetic exchange
interaction between neighboring spins. Numerical studies of this model
have been hampered, however, by the smallness of the RA strength $D_{R}$
compared to the strength of the exchange interaction $J$. Values
of $D_{R}\ll J$ would be typical for the structural disorder at the
atomic scale in amorphous ferromagnets. Scaling of $R_{f}$ as some
dimensionality-dependent power \citep{IM,CSS-1986} of $J/D_{R}$
makes the ferromagnetic correlation length $R_{f}$ large compared
to the interatomic distance $a$. As an example, magnetic state of
the system of $300\times300$ spins obtained by the minimization of
energy starting from random initial conditions (RIC) for $D_{R}/J=0.1$
is shown in Fig.\ \ref{Fig-magnetic_structure}. In a system of the
linear size $L$, the condition for the scaling $a\ll R_{f}\ll L$
at $D_{R}\ll J$ requires a very large $L$. This, in turn, requires
impractically large computing times. Computations performed in the
past \citep{Serota-1986,Dieny-PRB1990,DC-PRB1991,Fisch-1998,Fisch-2000,PCG-2015,GC-JPhys2022}
used $D_{R}\gg J$, $D_{R}\sim J$ or, at best, $D_{R}$ of the order
of $0.1J$. 

Past theoretical studies \citep{Saslow2018} of dynamical properties
of the RA magnets mostly focused on the experimentally observed \citep{Monod,Prejean,Alloul1980,Schultz,Gullikson}
broadening of the FMR line in the presence of strong magnetic field,
and on the localization of spin modes \citep{Fert,Levy,Henley1982,HS-1977,Saslow1982,Bruinsma1986,Serota1988,Ma-PRB1986,Zhang-PRB1993,Alvarez-PRL2013,Yu-AnnPhys2013,Nowak2015}
that had been reported in experiments on several magnetic systems
with static randomness \citep{Amaral-1993,Suran1-1997,Suran2-1997,Suran-1998,McMichael-PRL2003,Loubens-PRL2007,Du-PRB2014}.
More recently, it was demonstrated \citep{GC-PRB2021,GC-PRB2022,GC-PRB2023L,CG-PRB2023IP}
that amorphous and granular ferromagnets can be promising materials
for strong broadband microwave absorption. The latter has recently
been a focus of the growing number of research papers inspired by
applications for microwave shielding and stealth technology.

In Ref.\ \onlinecite{CG-RB2024} the scaling argument specific
to static randomness has been proposed that allows one to make conclusions
about the behavior of a large system with a weak disorder by studying
a smaller system with a strong disorder. The argument was based on
the scaling properties of the Hamiltonian of the RA model. In this
article we focus on the scaling of the equations describing the temporal
behavior of the RA system. Rescaling the Landau-Lifshitz equation
and using the fluctuation-dissipation theorem, we derive in terms
of the rescaled variables the expression for the frequency dependence
of the absorbed microwave power and show how the use of the dynamical
scaling drastically reduces the computing time.

The article is organized as follows. In Section \ref{Sec-statics}
we re-derive in a slightly different form the results of Ref.\ \onlinecite{CG-RB2024}
for the scaling of the static properties of the RA model. Scaling
of the dynamical equations and the results for the absorbed microwave
power are presented in Section \ref{Sec-dynamics}. Section \ref{Sec-discussion}
contains discussion of our results and conclusions.

\section{Statics}

\label{Sec-statics}

\subsection{Scaling of the RA model}

We consider a classical RA model of an amorphous ferromagnet on a
$d$-dimensional hypercubic lattice of spacing $a$ with the Hamiltonian
\begin{equation}
{\cal H}=-\frac{1}{2}\sum_{ij}J_{ij}{\bf s}_{i}\cdot{\bf s}_{j}-\frac{1}{2}D_{R}\sum_{i}({\bf n}_{i}\cdot{\bf s}_{i})^{2},\label{Hamiltonian}
\end{equation}
where $J_{ij}$ is the nearest-neighbor exchange interaction of strength
$J$, and ${\bf s}_{i}$ and ${\bf n}_{i}$ are the unit spin vector
and the unit vector of the local uniaxial magnetic anisotropy at the
position $i$ in the lattice. We assume that $D_{R}\ll J$, so that
the ferromagnetic correlation length $R_{f}$ is large compared to
$a$. In this case, instead of considering individual spins, one can
consider blocks of spins of size $b$ satisfying $a\ll b\ll R_{f}$
with $b$ being a multiple of $a$. In a system of $N$ spins, the
number of spins within the block, $n_{b}$, and the number of blocks,
$N_{b}$, are given by 
\begin{equation}
n_{b}=\left(\frac{b}{a}\right)^{d},\qquad N_{b}=\left(\frac{a}{b}\right)^{d}N.
\end{equation}

Since the directions of the lattice spins acquire a significant change
at a distance $R_{f}\gg a$, one can use the continuous approximation
for the energy (\ref{Hamiltonian}): 
\begin{equation}
{\cal H}=\frac{1}{2}\int\frac{d^{d}r}{a^{d}}\left[Ja^{2}(\nabla{\bf s})^{2}-D_{R}({\bf n}\cdot{\bf s})^{2}\right],\label{continuous}
\end{equation}
with ${\bf s}$ being the dimensionless spin density of unit length.
Discretizing the exchange part again with the lattice constant $b$
one obtains

\begin{equation}
\mathcal{H}_{ex}=-\frac{1}{2}\sum_{lk}J'_{lk}\mathbf{s}_{l}\cdot\mathbf{s}_{k},\label{Ham_ex_b}
\end{equation}
where now the summation runs over the blocks and $\mathbf{s}_{l}$
are block spins, also normalized by one. The new exchange constant
$J'$ can be obtained from the condition that the continuous forms
of these energies are the same:

\begin{equation}
{\cal H}_{ex}=\frac{1}{2}\int\frac{d^{d}r}{a^{d}}\left[Ja^{2}(\nabla{\bf s})^{2}\right]=\frac{1}{2}\int\frac{d^{d}r}{b^{d}}\left[J'b^{2}(\nabla{\bf s})^{2}\right].
\end{equation}
This yields
\begin{equation}
J'=\left(\frac{b}{a}\right)^{d-2}J.\label{J-prime}
\end{equation}

To rescale the random magnetic anisotropy, one must sum up all RA
terms within the spin blocks taking into account that all spins within
the block are collinear. This yields the rescaled anisotropy energy
given by 
\begin{equation}
{\cal H}_{RA}=-\frac{1}{2}D_{R}\sum_{l}t_{\alpha\beta,l}s_{l\alpha}s_{l\beta},\qquad t_{\alpha\beta,l}\equiv\sum_{i\in l}n_{l\alpha}n_{l\beta},\label{H-RA}
\end{equation}
where $t_{\alpha\beta,l}$ is a symmetric tensor defined by the summation
over the grain $l$. Its elements are of order $\sqrt{n_{b}}\sim(b/a)^{d/2}$.
Any such tensor can be brought to a diagonal form by choosing mutually
perpendicular principal coordinate axes. Consequently, the effective
RA of each block is a biaxial anisotropy in the coordinate axes specific
to that block. It can be rewritten as 
\begin{equation}
{\cal H}_{RA}=-\frac{D'_{R}}{2}\sum_{l}T_{\alpha\beta,l}s_{l\alpha}s_{l\beta},\quad T_{\alpha\beta,l}\equiv\left(\frac{a}{b}\right)^{d/2}\sum_{i\in l}n_{l\alpha}n_{l\beta}\label{H-RA-blocks}
\end{equation}
with the elements of the $T_{\alpha\beta,l}$ tensor being of order
unity and the effective anisotropy strength of the block given by
\begin{equation}
D'_{R}=\left(\frac{b}{a}\right)^{d/2}D_{R}.\label{DR-prime}
\end{equation}
The full spin-block Hamiltonian becomes 
\begin{equation}
{\cal H}=-\frac{1}{2}\sum_{lk}J'_{lk}{\bf s}_{l}\cdot{\bf s}_{k}-\frac{1}{2}D'_{R}\sum_{l}T_{\alpha\beta,l}{\bf s}_{l\alpha}{\bf s}_{l\beta}.\label{Block-Hamiltonian}
\end{equation}

As the spin-block model is equivalent to the original atomic-spin
model the ferromagnetic correlation length for the two models should
be the same. Thus, up to a dimensionality-dependent factor $k_{d}$,
it can be obtained by equating the two expressions for $R_{f}$: 
\begin{equation}
R_{f}=k_{d}\left(\frac{J}{D_{R}}\right)^{p}a=k_{d}\left(\frac{J'}{D'_{R}}\right)^{p}b,
\end{equation}
where $p$ is the power to be determined. Substituting here $J'$
and $D'_{R}$ given be the equations (\ref{J-prime}) and (\ref{DR-prime}),
we obtain 
\begin{equation}
R_{f}=k_{d}\left(\frac{J}{D_{R}}\right)^{p}a=k_{d}\left(\frac{J}{D_{R}}\right)^{p}a\left(\frac{b}{a}\right)^{(d/2-2)p+1},
\end{equation}
which produces a well-known result 
\begin{equation}
p=\frac{2}{4-d}
\end{equation}
and thus \citep{IM,CSS-1986} 
\begin{equation}
R_{f}=k_{d}\left(\frac{J}{D_{R}}\right)^{\frac{2}{4-d}}a.\label{Rf_IM}
\end{equation}
This result suggests that in less than four dimensions, the long-range
ferromagnetic order is destroyed by the RA no matter how weak because
$R_{f}$ is finite. This result should be taken with caution for two
reasons. First, the finite result for $R_{f}$ was obtained under
the assumption that it is a power of $J/D_{R}$. Second, the complex
energy landscape of the RA system contains barriers that prevent it
from reaching the lowest-energy state during typical times of real
and numerical experiments. Such a system exhibits magnetic hysteresis
\citep{PCG-2015} with the area of the hysteresis loop increasing
on lowering temperature. For the $RA$ model $R_{f}$ was never computed
analytically.

\subsection{Ferromagnetic correlation length}

The existence of a finite correlation length $R_{f}\gg a$ creates
an extended short-range order in the RA magnet. Due to statistical
fluctuations, in the absence of a long-range order, a finite system
acquires a finite magnetization per spin that can be expressed as
\citep{GC-JPhys2022} 
\begin{equation}
m^{2}=\frac{1}{N}\sum_{j}{\bf s}_{i}\cdot{\bf s}_{j}=\frac{1}{N}\int_{0}^{\infty}\frac{d^{d}r}{a^{d}}C(r),
\end{equation}
where $C(r)$ is the spin-spin correlation function. This result can
be averaged over different realizations of the states of the system.
As the analytical form of $C(r)$ is unknown, we resort to approximating
it as $C(r)=\exp[-(r/R_{f})^{p}]$. Then in 2D one obtains
\begin{equation}
m^{2}=K_{p}\frac{\pi R_{f}^{2}}{Na^{2}}\quad\Longrightarrow\quad\frac{R_{f}}{a}=m\sqrt{\frac{N}{\pi K_{p}}},\label{Rf_2D_via_m}
\end{equation}
where $K_{1}=2$ and $K_{2}=1$. The result in 3D reads
\begin{equation}
m^{2}=K_{p}\frac{\pi R_{f}^{3}}{Na^{3}}\quad\Longrightarrow\quad\frac{R_{f}}{a}=\left(\frac{m^{2}N}{\pi K_{p}}\right)^{1/3},\label{Rf_3D_via_m}
\end{equation}
where $K_{1}=8$ and $K_{2}=\sqrt{\pi}$. 

\begin{figure}
\begin{centering}
\includegraphics[width=8cm]{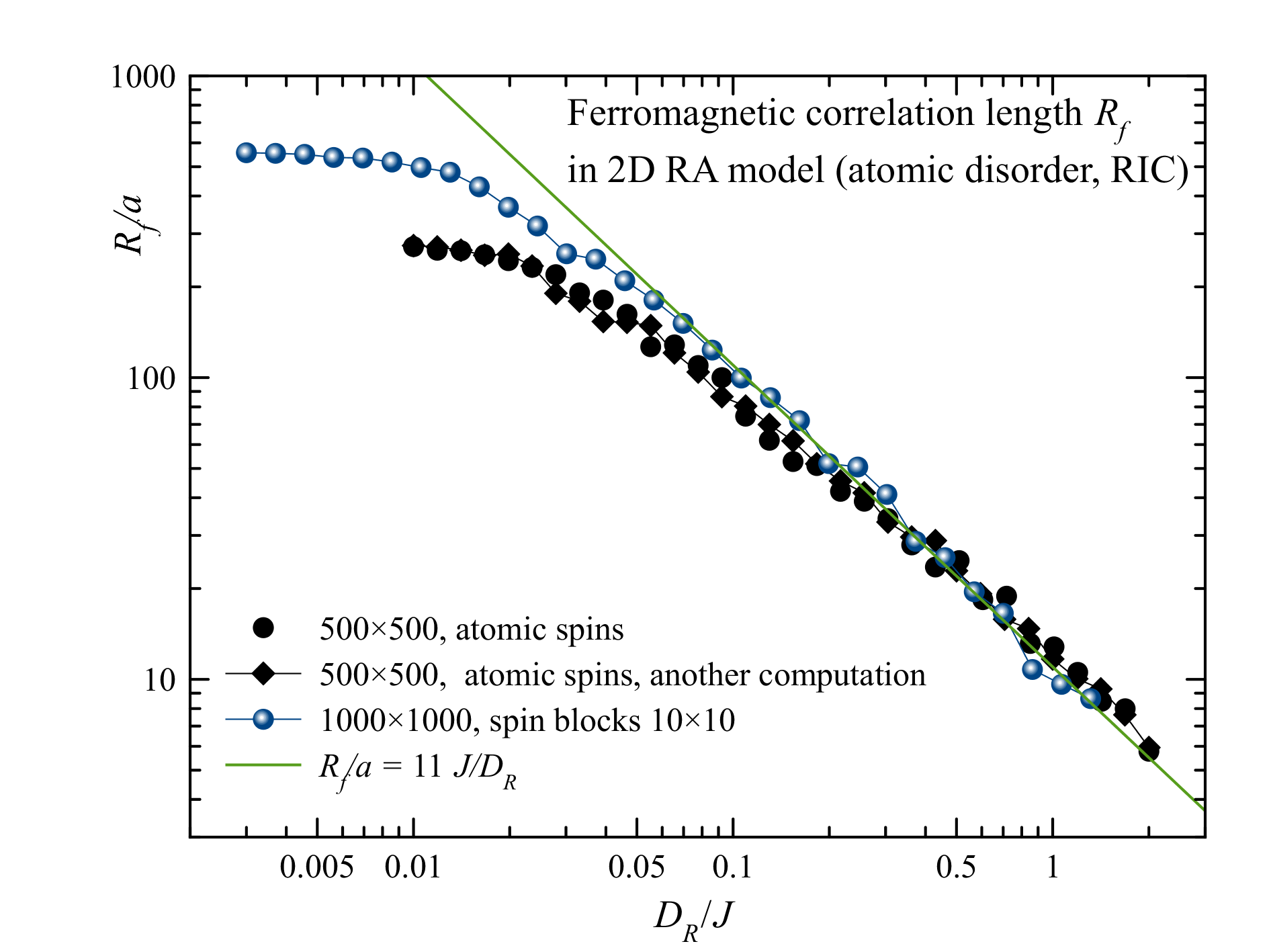} \includegraphics[width=8cm]{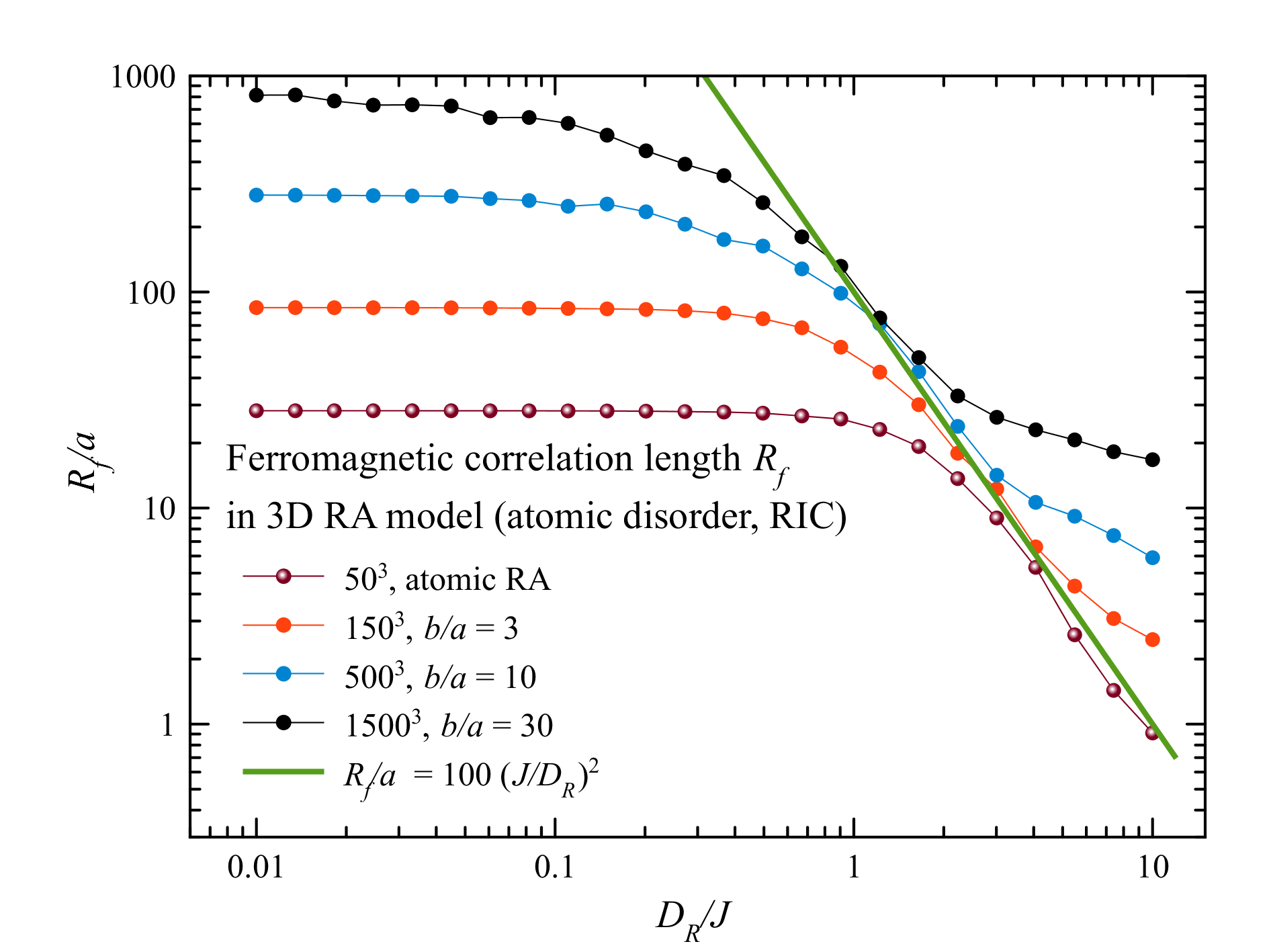}
\par\end{centering}
\caption{Ferromagnetic correlation radius $R_{f}$ in the RA model with atomic
disorder, computed in terms of atomic spins and spin blocks. Upper
panel: 2D model, $L=500,1000$. Lower panel: 3D model, $L=50,150,500,1500$.}

\label{Fig-Rf}
\end{figure}

Applicability of the formulas above requires $a\ll R_{f}\ll L$, where
$L$ is the linear size of the system. For $D_{R}/J\ll1$ a very large
$L$ is needed because $R_{f}$ is large. This makes computations
difficult. However, if one uses spin blocks of size $b$ satisfying
$a\ll b\ll R_{f}$, with a spin of the block still normalized to one,
the average magnetization is 
\begin{equation}
m^{2}=\frac{1}{N_{b}}\sum_{j}{\bf s}_{i}\cdot{\bf s}_{j}=\frac{1}{N_{b}}\int_{0}^{\infty}\frac{d^{d}r}{b^{d}}C(r).
\end{equation}
This results in $R_{f}/b=m\sqrt{N_{b}/\left(\pi K_{p}\right)}$ in
2D and $R_{f}/b=\left[m^{2}N_{b}/\left(\pi K_{p}\right)\right]^{1/3}$
in 3D, which allows one to obtain the dependence of $R_{f}$ on $D_{R}$
in a wide range of $D_{R}$ by changing the size of the spin block
$b$.

The states of the RA system corresponding to a local energy minimum
were found by the energy minimization starting from the random initial
condition (RIC) for the system with free boundary conditions. The
numerical method uses spin flips over the effective field (overrelaxation),
${\bf s}_{i}\to2({\bf s}_{i}\cdot{\bf H}_{{\rm eff},i}){\bf H}_{{\rm eff},i}/H_{{\rm eff},i}^{2}-{\bf s}_{i}$.
While for different-site interactions, such as the exchange, overrelaxation
conserves the energy, for our model with the single-site anisotropy
it leads to the energy lowering, see Eq. (3) of Ref. \citep{GC-PRB2023L}.
This method proves to be very efficient and is much better than the
straightforward aligning the spins with their effective fields.

The ferromagnetic correlation length $R_{f}$ in the 2D RA model with
atomic disorder computed in terms of atomic spins and spin blocks
and averaged over 32 realization of the random magnetic state at $T=0$
is shown in the upper panel of Fig. \ref{Fig-Rf}. There is a good
agreement between the two types of computation, as well as with the
scaling prediction, Eq. (\ref{Rf_IM}), in which the numerical factor
is estimated as $k_{2}\simeq11$.

The 3D case is challenging because $R_{f}$ depends quadratically
on $J/D_{R}$ and becomes very large for $J/D_{R}\apprle1$. As the
two inequalities, $a,b\ll R_{f}\ll L$, have to be satisfied and thus
a large $L$ is required, in 3D this leads to an exceedingly large
number of spins in the system that makes numerical calculation difficult.
Apriori, the scaling formula, Eq. \ref{Rf_IM}, has to be valid within
at least one decade of $J/D_{R}$ to be confirmed. This requires,
minimally, $10\lesssim R_{f}/a\lesssim100$ and thus $L/a\simeq1000$
that amounts to $10^{9}$ spins in the system. We computed $R_{f}$
in the system of $50^{3}=125000$ atomic spins or spin blocks with
linear sizes $b/a=3,10,30$. The results in the lower panel of Fig.
\ref{Fig-Rf} obtained by averaging over 32 realizations show that
for these insufficiently large systems Eq. \ref{Rf_IM} is valid in
a rather narrow range of $D_{R}$. At larger $D_{R}$, the condition
$b\ll R_{f}$ is violated and the computed values of $R_{f}$ are
overestimated. At smaller $D_{R}$, the condition $R_{f}\ll L$ is
violated and the values of $R_{f}$ are underestimated (the computed
$R_{f}$ saturates at approximately $L/2$). Still, the assembly of
these regions for different $b/a$ reproduces the scaling formula,
Eq. \ref{Rf_IM} and allows to find the coefficient $k_{3}\simeq100$. 

It has to be mentioned that the state of the RA system after the energy
minimization depends on the initial state. Disordered states with
finite $R_{f}$ are obtained here starting from the RIC. To the contrary,
starting from the collinear initial condition leads to only partially
disordered states with a significant residual magnetization, $m\sim1$.
These differences are due to the energy barriers in the RA system.

\subsection{Correlated disorder}

In practice, one often deals with a correlated rather than atomic-scale
disorder, when the magnetic anisotropy axes are aligned on a scale
$R_{a}>a$. This can be due to the short-range structural order in
an amorphous ferromagnet or can occur in a magnet sintered of crystalline
ferromagnetic grains of size $R_{a}$. The correlated disorder with
$b=R_{a}$ does not change the expression for $J'$ in Eq.\ (\ref{J-prime})
but it changes $D'_{R}$ in Eq.\ (\ref{DR-prime}) to $D'_{R}=(R_{a}/a)^{d}D_{R}$.
The ferromagnetic correlation length changes accordingly \citep{CG-RB2024}:
\begin{equation}
\frac{R_{f}}{a}=k_{d}\left(\frac{J}{D_{R}}\right)^{2/(4-d)}\left(\frac{a}{R_{a}}\right)^{d/(4-d)}.\label{Ra}
\end{equation}
It decreases with increasing $R_{a}$ as $1/R_{a}$ in 2D, and much
more rapidly, as $1/R_{a}^{3}$, in 3D. This allows one to satisfy
the condition $R_{f}\ll L$ in a system of a smaller size, which simplifies
the numerical work. Notice that Eq.\ (\ref{Ra}) is valid for $R_{f}\gg R_{a}$.
When $R_{f}$ given by this equation becomes smaller than $R_{a}$,
one must set $R_{f}=R_{a}$, which corresponds to independently ordered
ferromagnetic grains.

\section{Dynamics}

\label{Sec-dynamics}

\subsection{Scaling of the Landau-Lifshitz equation}

We now turn to the rescaling of the dynamical equations describing
the RA system, which has not been done in the past. The equations
of motion for individual spins in the absence of the phenomenological
damping are 
\begin{equation}
\hbar\frac{\partial{\bf s}_{i}}{\partial t}={\bf s}_{i}\times{\bf H}_{{\rm eff},{i}},\label{LL}
\end{equation}
where ${\bf H}_{{\rm eff},{i}}$ is the effective field acting on
the spin $i$. It is given by 
\begin{equation}
{\bf H}_{{\rm eff},{i}}=-\frac{\partial{\cal H}}{\partial{\bf s}_{i}}=\sum_{j}J_{ij}{\bf s}_{j}+D_{R}({\bf n}_{i}\cdot{\bf s}_{i}){\bf n}_{i}.\label{H-eff}
\end{equation}

Within correlated regions of size $R_{f}$ the lattice spins are precessing
synchronously. The spins of the blocks of size $b\ll R_{f}$ are precessing
at the same rate as the lattice spins. Their equations of motion are
\begin{equation}
\hbar\frac{\partial{\bf s}_{l}}{\partial t}={\bf s}_{l}\times{\bf \bar{H}}_{{\rm eff},{l}},\label{LL-blocks}
\end{equation}
where ${\bf \bar{H}}_{{\rm eff},{l}}$ is the effective field per
spin for the $l$-th block given by 
\begin{equation}
{\bf \bar{H}}_{{\rm eff},{l}}=-\frac{1}{n_{b}}\frac{\partial{\cal H}}{\partial{\bf s}_{l}}=\frac{1}{n_{b}}\left[\sum_{k}J'_{lk}{\bf s}_{k}+D'_{R}\mathbb{T}_{l}\cdot{\bf s}_{l}\right],\label{H-eff-block}
\end{equation}
where $(\mathbb{T}_{l}\cdot{\bf s}_{l})_{\alpha}=T_{\alpha\beta,l}s_{l\beta}$.
It can be written as 
\begin{equation}
{\bf \bar{H}}_{{\rm eff},{l}}=\sum_{k}\bar{J}_{lk}{\bf s}_{k}+\bar{D}_{R}\mathbb{T}_{l}\cdot{\bf s}_{l}\label{H-eff-block-bar}
\end{equation}
with 
\begin{equation}
\bar{J}=\left(\frac{a}{b}\right)^{2}J,\qquad\bar{D}_{R}=\left(\frac{a}{b}\right)^{d/2}D_{R}.\label{bar-JD}
\end{equation}
Here $\bar{D}_{R}$ is the effective magnetic anisotropy per spin
of the block. Its use is justified by the fact that it determines
the FMR frequency in a conventional ferromagnet.

One advantage of the above rescaling of the dynamical equations is
that the model of spin blocks has now fewer variables than the original
lattice-spin model by a factor $(a/b)^{d}$. Also, since 
\begin{equation}
\frac{\bar{J}}{\bar{D}_{R}}=\left(\frac{a}{b}\right)^{(4-d)/2}\frac{J}{D_{R}},
\end{equation}
the exchange/anisotropy ratio is now reduced by a factor $(a/b)^{(4-d)/2}$,
which represents another advantage for the numerical work. In the
numerical integration of the equations of motion, the typical integration
step for the model of lattice spins is about $\Delta t=0.1\hbar/(J+D_{R})$.
Since $D_{R}\ll J$, it is determined by the exchange. For the spin-block
model the integration step $\Delta t=0.1\hbar/(\bar{J}+\bar{D}_{R})$
is much larger. This greatly speeds up the computation.

For a correlated disorder corresponding to the ferromagnetically ordered
grains of size $b=R_{a}$, and effective exchange and anisotropy given
by $J'=\left(b/a\right)^{d-2}J$ and $D'_{R}=(R_{a}/a)^{d}D_{R}$,
one has the same expression for $\bar{J}$ as in Eq.\ (\ref{bar-JD})
but $\bar{D}_{R}=D_{R}$. Now $\bar{J}/\bar{D}_{R}=(a/R_{a})^{2}(J/D_{R})$,
which provides even greater advantage for the numerical work in two
and three dimensions.

\subsection{Microwave absorption}

In a system with a complex energy landscape, such as the RA system,
achieving a full equilibration over all the energy minima at low temperatures
is practically impossible. Instead, thermal equilibrium is achieved
inside the local energy minimum that represents a certain point on
the hysteresis curve. To compute the microwave absorption at any temperature
$T$, one can use the fluctuation-dissipation theorem (FDT) which
provides the following expression for the absorbed power of microwaves
of frequency $\omega$ and the amplitude of the magnetic field $h_{0}$:
\begin{equation}
P=\frac{\omega^{2}h_{0}^{2}}{6Nk_{B}T}\mathrm{Re}\left[\int_{0}^{\infty}dte^{i\omega t}\langle{\bf S}(0)\cdot{\bf S}(t)\rangle\right],
\end{equation}
where ${\bf S}(t)=\sum_{i}{\bf s}_{i}$ is the total spin of the system.

Expressing ${\bf S}$ via spins of the spin blocks as 
\begin{equation}
{\bf S}=n_{b}\sum_{l}{\bf s}_{l}=n_{b}N_{b}{\bf m}_{b}=N{\bf m}_{b},
\end{equation}
where 
\begin{equation}
{\bf m}_{b}\equiv\frac{1}{N_{b}}\sum_{l}{\bf s}_{l}
\end{equation}
is the average spin of the block, one can re-write the absorbed power
per spin as 
\begin{equation}
{P}=\frac{\omega^{2}Nh_{0}^{2}}{6k_{B}T}\mathrm{Re}\left[\int_{0}^{\infty}dte^{i\omega t}\langle{\bf m}_{b}(0)\cdot{\bf m}_{b}(t)\rangle\right],\label{power}
\end{equation}

\begin{center}
\begin{figure}[h]
\begin{centering}
\includegraphics[width=8cm]{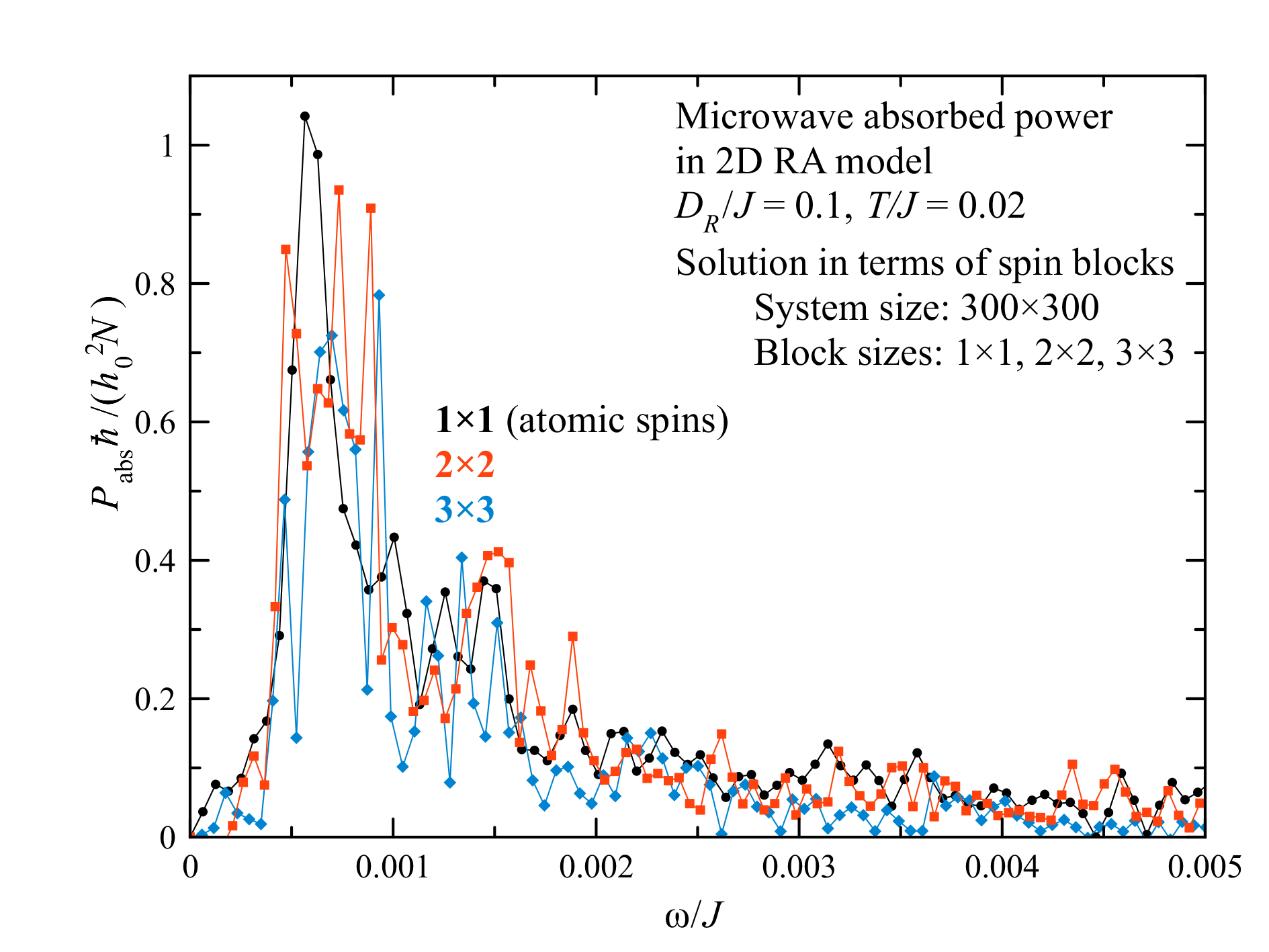} 
\par\end{centering}
\centering{}\includegraphics[width=8cm]{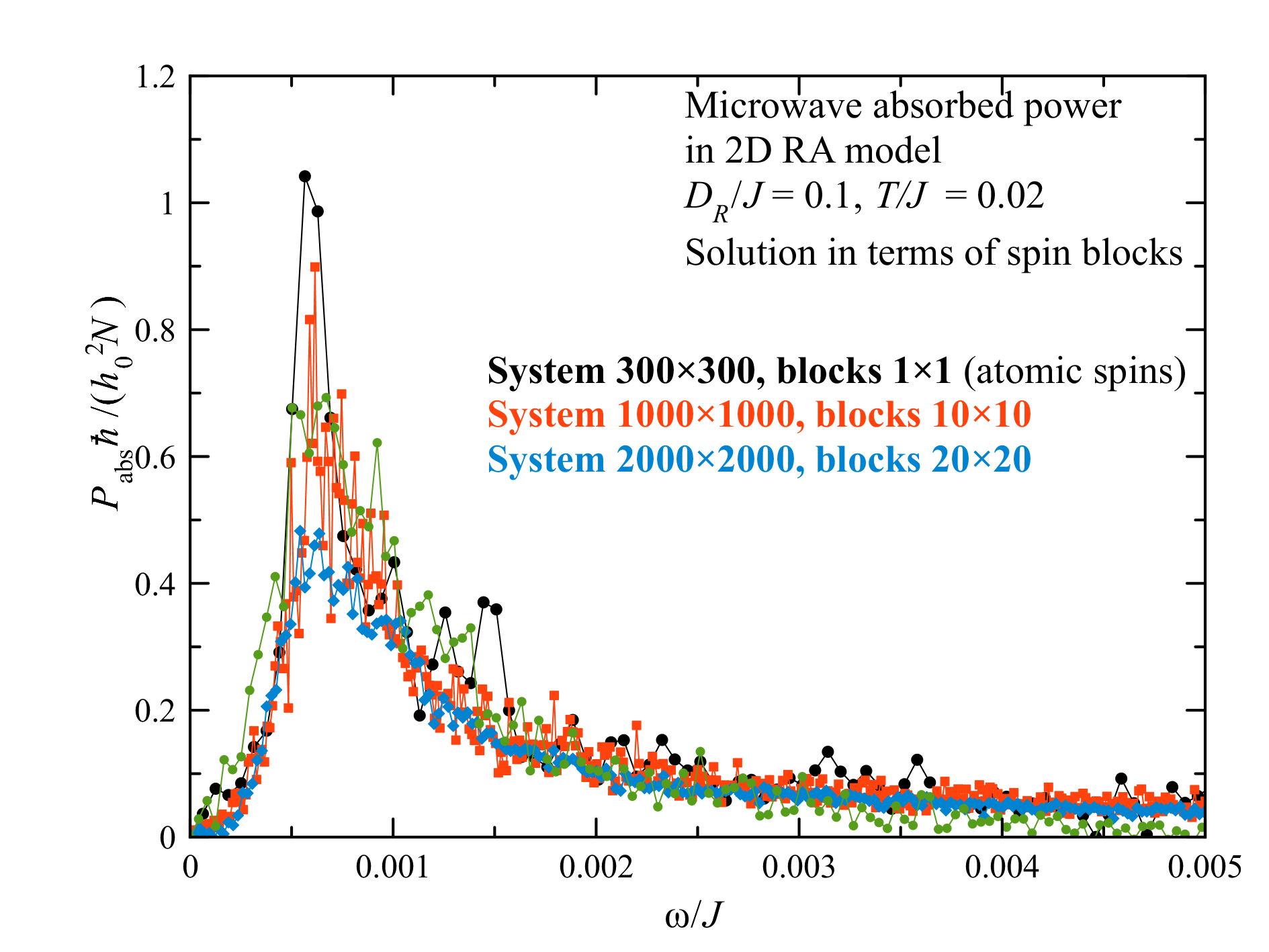}\caption{Frequency dependence of the absorbed power for original and rescaled
2D RA models at $D_{R}/J=0.1$ and low temperature. Upper panel: The
same $300\times300$ system solved in terms of the original atomic
spins and spin blocks. Lower panel: Systems of different sizes solved
in terms of atomic spins and spin blocks.}
\label{Fig-Absorbed_power} 
\end{figure}
\par\end{center}

In the numerical solution of the rescaled problem, first, a thermal
state of spin blocks is created with the Monte Carlo equilibration.
At this stage, one uses the energy-rescaled model, Eq.\ (\ref{Block-Hamiltonian}).
Then one integrates the equation of motion for the system of blocks,
Eqs.\ (\ref{LL-blocks}) and (\ref{H-eff-block-bar}). The comparison
of the results obtained by the direct numerical solution of the 2D
spin-lattice model and the solution of the spin-block model for $D_{R}/J=0.1$
are shown in Fig.\ \ref{Fig-Absorbed_power}. 

In films of thickness $h\ll R_{f}$, the spins along the $z$-axis
normal to the film have the same orientation at $T\ll J$. This allows
for a simple generalization of the above scaling. Introducing blocks
of volume $b\times b\times h$, one obtains Eqs.\ (\ref{H-RA}) and
(\ref{H-RA-blocks}) in which the summation goes over these 3D blocks
while the exchange part of the Hamiltonian is simply multiplied by
the number of the atomic layers $N_{z}$. Due to the additional summation
of the RA along the $z$-axis, the effective RA field per spin of
the block further decreases as compared to a 2D model. Consequently,
the microwave absorption maximum shifts to lower frequencies. The
absorption power per spin can be obtained by using Eq.\ (\ref{power})
with the additional factor $N_{z}$, while the initial state of the
system should be prepared by Monte Carlo with the spin-block Hamiltonian
obtained by the summation of the exchange and RA terms over the 3D
blocks.

\section{Discussion}

\label{Sec-discussion}

For decades, the random-anisotropy model has been successfully applied
to interpreting experimental results on amorphous and sintered ferromagnets.
Its key ideas are based on the Imry-Ma argument \citep{IM}. While
the theory allows one to estimate the range of the resulting short-range
order, the numerical studies of this phenomenon have always been hampered
by the necessity to consider a system of a prohibitively large number
of spins. Scaling of the static properties of the random-anisotropy
model recently proposed by us \citep{CG-RB2024} has alleviated this
difficulty. In this article, we extended our scaling ideas to the
dynamical properties of the random-anisotropy magnet studied in terms
of the Landau-Lifshitz equations for individual spins. 

Before studying the dynamical scaling, we re-derived our scaling relations
for the static properties of the random-anisotropy magnet by an alternative
method. It allowed us to obtain the Imry-Ma relations based entirely
upon the scaling argument without commonly used energy minimization.
The advantage of the scaling is demonstrated by the excellent agreement
of the theory with numerical results on the parameter dependence of
the ferromagnetic correlation length, which had been previously unavailable.
They have been obtained by switching to blocks of spins interacting
via the rescaled exchange and anisotropy constants. 

Moving to the dynamics of the spins described by the Landau-Lifshitz
equations, we have implemented a dynamical scaling of the random-anisotropy
model in terms of spin blocks. By utilizing a fewer number of variables
in a large system containing up to one million spins, we have demonstrated
that the rescaled model permits the use of a much larger integration
step in the numerical code. This greatly speeds up computation. 

The proposed dynamical scaling has been used to obtain the expression
for the frequency dependence of the absorbed microwave power by the
random-anisotropy magnet in terms of spin blocks. To test our scaling
method we computed the absorbed power in the rescaled model with the
spin blocks and compared it with the power computed for the original
model dealing with a much greater number of individual atomic spins.
Up to the numerical noise, the results are identical, but they require
orders of magnitude shorter computer time for the rescaled model. 

We have shown how our scaling method can be extended to thin films
and magnetic systems sintered of nanograins with magnetic anisotropy
correlated within the grain. While we stick to the random-anisotropy
model, the ideas presented here can be utilized to obtain computational
advantage for other systems with quenched randomness, such as, e.g.,
pinned charge-density waves and flux lattices in superconductors. 

\section*{ACKNOWLEDGEMENT}

This work has been supported by Grant No. FA9550-24-1-0090 funded
by the Air Force Office of Scientific Research.

\end{document}